\documentclass[12pt]{article}
\usepackage[italian,english]{babel}
\usepackage{graphicx,lscape}
\usepackage{verbatim}
\usepackage{latexsym}
\usepackage{amsmath,amsfonts,amssymb,amsthm,wasysym}
\def\bseq{\begin{subequation}}  
\def\eseq{\end{subequation}}
\def\bsea{\begin{subeqnarray}}  
\def\esea{\end{subeqnarray}}

\newcommand{\bbox}{\lower.2ex\hbox{$\Box$}}

\newcommand{\beq}{\begin{equation}}
\newcommand{\eeq}{\end{equation}}
\newcommand{\bea}{\begin{eqnarray}}
\newcommand{\eea}{\end{eqnarray}}
\newcommand{\ena}{\end{eqnarray}}

\newcommand {\non}{\nonumber}

\renewcommand{\[}{\left[}

\newcommand{\be}{\begin{equation}}
\newcommand{\ee}{\end{equation}}

\begin{document}
\begin{titlepage}
{\hbox to\hsize{June 2007 \hfill}}
\begin{center}
\vglue .06in
\vskip 40pt

{\Large\bf Meta-stable  
$A_n$ 
quiver gauge theories
%
} 
\\[.8in]
{\large\bf A. Amariti\footnote{antonio.amariti@unimib.it},
L. Girardello\footnote{luciano.girardello@mib.infn.it} and
A. Mariotti\footnote{alberto.mariotti@mib.infn.it}
}
\\[.4in]
{\it Dipartimento di Fisica, Universit\`a degli Studi di
Milano-Bicocca\\ 
and INFN, Sezione di Milano-Bicocca, piazza della Scienza 3, 
I 20126 Milano, Italy}
\\[.2in]

\vskip 10pt

{\bf ABSTRACT}\\[.3in]
\end{center}
We study metastable dynamical breaking of supersymmetry in $A_n$ 
quiver gauge theories.
We present a general analysis and 
criteria for
the perturbative existence of metastable vacua 
in quivers of any length.
Different mechanisms of gauge mediation can be realized. 

\vskip 10pt
${~~~}$ \newline
\\[2in]  
\vspace{8cm}

\end{titlepage}

\section{Introduction}

 The existence of long living metastable vacua 
\cite{ISS,rattazzi}
seems by now a rather
generic phenomenon in large classes of supersymmetric gauge
theories \cite{Ooguri1,Franco1,Forste}.
It provides an attractive way for dynamical breaking of
supersymmetry and the interest in these theories has been enhanced by
the possibilities of their embedding in supergravity and string theory 
\cite{Ooguri2}
and of their use 
\cite{Dine1,Kitano,Kawano,Murayama1}
in gauge mediation mechanisms
\cite{mediation}.

Metastability is a low energy phenomenon for UV free theories and in
general the key ingredient which makes  a perturbative
analysis possible is 
Seiberg duality to IR free theories described in terms of
macroscopic fields
\cite{Seibergd}.

An interesting set of theories in which to study metastability \`a la ISS 
\cite{ISS} 
is the ADE class of quiver gauge theories
\cite{Cachazo1,Oh}.

These theories can be derived in type $IIB$ string theory
from $D5$-branes partially wrapping 2-cycles of 
non compact
Calabi-Yau threefolds. These manifolds are ADE-fold
geometries fibered over a plane, and the 2-cycles are
blown up $S_i^2$ in one
to one correspondence with the simple roots
of ADE.

In this paper we investigate metastability in $A_n$  
$\mathcal{N}=2$ (non affine) quiver gauge theories
deformed to $\mathcal{N}=1$ by superpotential terms in the
adjoint fields.
In the presence of many gauge
groups we have, in principle, a large number of dualization
choices. 

In \cite{Ooguri1,Kitano,Kawano} $A_2,A_3,A_4$ quivers have been studied
dualizing only 
one node in the quiver, 
where dynamical supersymmetry 
breaking occurs.

Here we consider $A_n$ theories with arbitrary $n$, 
where several Seiberg dualities take place.
In particular 
we will explore theories obtained by dualizing alternate nodes.
This leads to a low energy description in terms of only
magnetic fields. 

In the duality process the dualized groups are treated as
genuine gauge groups whereas the other ones have to be weakly coupled
at low energy, so that they act as flavour groups i.e. global symmetries.
The procedure depends on the interplay of the RG flows of the dualized
and of the non dualized gauge groups and is governed by the associated
beta-functions. This translates into inequalities among the ranks
of the gauge groups and in hierarchies among the strong coupling scales.

The paper is organized as follows.
In section 2 we describe the $\mathcal N = 2$
quiver gauge theories, explicitly 
broken to $\mathcal N = 1$ by superpotential
terms. After the integration of the massive adjoint fields,
we give the general form of the superpotential.
In section 3 we investigate Seiberg
duality on the alternate nodes of the quiver. The general
theory obtained with this procedure on an $A_n$ is 
expressed in terms of only magnetic fields.
In section 4 we 
consider the simplest case, i.e. $A_3$ quiver,
showing that it possesses
long
living metastable vacua \`a la ISS. The analysis 
is done neglecting the gauge contributions of the odd 
nodes, which are treated as flavour symmetries. 
This last approximation is justified in section 5,
where an analysis of the running of the couplings 
has been performed.
The general result, metastability in an $A_n$ quiver theory,
is explained in section 6, giving an explicit example.
In section 7 we comment on the possible ways of enforcing
gauge mediation of supersymmetry breaking.
Appendix A explains how to find the metastable vacua upon changing
the masses of the quarks in the electric description.
Appendix B provides details in the analysis on the 
running of the gauge couplings of section 5.
Appendix C adds to section 6, giving all
the possible choices of $A_5$ which 
show metastable vacua.

\section{$A_n$ quiver gauge theories 
with massive adjoint fields}\label{section2}
We consider a $\mathcal{N}=2$ (non affine) $A_n$ quiver gauge theory,
deformed to $\mathcal{N}=1$ by superpotential terms 
in the adjoint fields. 
The theory is associated with a Dynkin diagram where
each node is a $U(N_i)$ gauge group. 
\begin{center}
\begin{tabular}{c}
    \includegraphics[width=15cm]{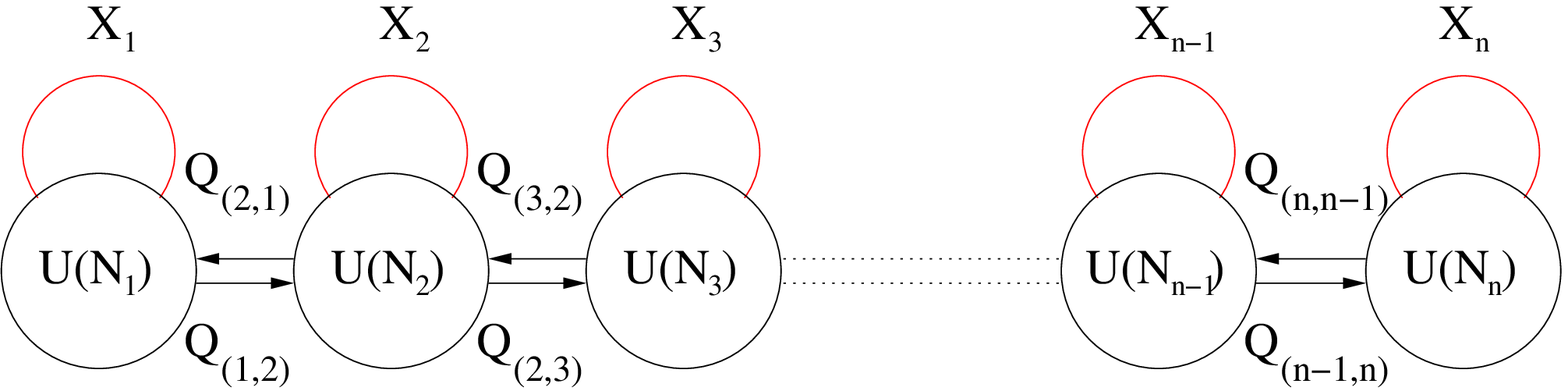}\\
\end{tabular}
\end{center}
The arrows connecting
two nodes represent fields 
$Q_{i,i+1}, Q_{i+1,i}$
in the fundamental of the incoming node and
anti fundamental of the out-coming node. 
The adjoint fields $X_i$
refer to the $i$-th gauge group. 

The gauge group 
of the whole theory is the product
$\prod_{i=1}^n U(N_i)$. 
We call $\Lambda_i$ the strong coupling scale 
of each gauge group.

The $\mathcal{N}=1$ superpotential is 
\begin{equation} \label{SpotIn}
W=\sum_{i=1}^{n} W_i(X_i) + \sum_{i,j} s_{i,j}(Q_{i,j})_{\alpha}^{\beta} 
(X_{j})_{\beta}^{\gamma}(Q_{j,i})_{\gamma}^{\alpha}
\end{equation}
where $s_{i,j}$ is an antisymmetric matrix, with $|s_{i,j}|=1$.
The Latin labels run on the different nodes of the $A_n$ quivers, 
the Greek labels runs on the ranks of the groups of each site.
In the case of $A_n$ theories the only non zero terms are $s_{i,i+1}$
and $s_{i,i-1}$.
The superpotentials for the adjoint fields $W_i(X_i)$
break supersymmetry to $\mathcal{N}=1$. 

We choose these
superpotentials 
to be
\begin{equation} \label{aggiunte}
W_i(X_i) = \lambda_i Tr{X_i} + \frac{m_i}{2}Tr{X_i^2}
\end{equation}
As a consequence the adjoint fields are all massive.
We consider the limit where the adjoint fields 
are so heavy that they can be integrated out, 
and we study the theory below the scale of their  
masses.

Integrating out these fields we obtain the 
 effective superpotential describing the $A_n$ theory
(traces on the gauge groups are always implied).
\begin{eqnarray} \label{SpotNoAdj}
W&=&\sum_{i=1}^{n-1} \left( \left( \frac{\lambda_{i+1}}{m_{i+1}}-
\frac{\lambda_i}{m_i}
\right) Q_{i,i+1}Q_{i+1,i} - \frac{1}{2} \left(\frac{1}{m_i}+
\frac{1}{m_{i+1}} \right) (Q_{i,i+1}Q_{i+1,i})^2 \right) \nonumber \\
&+& \sum_{i=2}^{n-1}\frac{1}{m_i}Q_{i-1,i}Q_{i,i+1}Q_{i+1,i}Q_{i,i-1}
\end{eqnarray}

A final important remark is that for the $A_n$ theories the $D$-term
equations of motion can be 
decoupled and simultaneously diagonalized \cite{csaki}.

\section{Seiberg duality on the even nodes}\label{alternate}
We investigate 
the low energy dynamics of the gauge groups
of the Dynkin diagram, governed 
by the ranks 
and by the hierarchy between the 
strong coupling scales of each node.
We work in the regime where 
the even nodes develop strong dynamics
and have to be 
Seiberg dualized. 

We set all the strong coupling scales of
the even nodes to be equal $\Lambda_{2i} \equiv \Lambda_G$
and we require the odd nodes to be less coupled at this scale.
We impose the following window for the ranks of the nodes
\be
\label{magnwind}
N_{2i}+1 \leq N_{2i-1}+N_{2i+1}< \frac{3}{2} N_{2i} 
\qquad i=1,\dots,\frac{n-1}{2}
\ee
We take $n$ odd, 
the even case can be included setting to zero one of the 
ranks of the extremal nodes.

Along the flow toward the IR, we have to change 
the description at the scale $\Lambda_G$ performing
Seiberg duality on the even nodes.
The even nodes are treated as gauge groups, whereas the
odd nodes are treated as flavours. We will discuss 
the consistency of this description in section 
\ref{BETAF}.

It is convenient to list the elementary fields of
the dualized theory, i.e. the electric gauge singlets and the
new magnetic quarks.
\small{
\begin{center}
\begin{tabular}{c|c|c|c}
&$U(N_{2i-1})$  &$U(\widetilde N_{2i})$ &$U(N_{2i+1})$ \\
\hline
$M_{2i+1,2i-1}$ & $N_{2i-1}$ &$1$ &
$\bar N_{2i+1}$  \\
\hline
$M_{2i+1,2i+1}$ & $1$ &$1$ & Bifund.  \\
\hline
$M_{2i-1,2i-1}$ & Bifund. &$1$ &$1$  \\
\hline
$M_{2i-1,2i+1}$ & $\bar N_{2i-1}$ &$1$ 
&$N_{2i+1}$  \\
\hline
$q_{2i-1,2i}$&$N_{2i-1}$& 
$\begin{array}{c}
\vspace{-4.4mm}
\\
\overline {\widetilde  N}_{2i}
\end{array}
$
&1
\\
\hline
$q_{2i,2i-1}$&$\bar N_{2i-1}$& 
$ 
\begin{array}{c}
\vspace{-4.4mm}
\\
\widetilde N_{2i}
\end{array}
$&$1$\\
\hline
$\phantom{\frac{1}{1}}$
$q_{2i,2i+1}$&1&
$\begin{array}{c}
\vspace{-4.4mm}
\\
\widetilde N_{2i}
\end{array}
$& $\bar N_{2i+1}$\\
\hline
$q_{2i+1,2i}$
&
1
&
$
\begin{array}{c}
\vspace{-4.4mm}
\\
\overline{\widetilde N}_{2i}
\end{array}
$
& $ N_{2i+1}$\\
\end{tabular}
\end{center}}
The mesons are proportional to the original electric
variables: $M_{2i+k,2i+j} \sim Q_{2i+k,2i}Q_{2i,2i+j}$.
The even magnetic groups have ranks $\widetilde N_{2i}=N_{2i+1}+N_{2i-1}-N_{2i}$.
The superpotential in
the new magnetic variables results 
\begin{eqnarray} \label{magnspot}
W &=& h  M_{2i+k,2i+j}^{(2i)}q_{2i+j,2i}q_{2i,2i+k}+h \mu_{2i+k, (2i)}^2  
M_{2i+k,2i+k}^{(2i)} +\\
&+&h m  M_{2i+1,2i+1}^{(2i)}M_{2i+1,2i+1}^{(2i+2)}
+ h m \left( M_{2i+k,2i+k}^{(2i)}\right)^2
+h m M_{2i-1,2i+1}^{(2i)}M_{2i+1,2i-1}^{(2i)} \non
\end{eqnarray}
where the index i runs from $1$ to $\frac{n-1}{2}$, 
and k and j are $+1$ or $-1$. 
The upper index $(2i)$
of the mesons indicates which site the meson refers to: it is
necessary because some mesons have the same flavor indexes, but they
are summed on different gauge groups, so they have to be
labeled differently. 
We denote with $h m_i$ the meson masses, related 
to the quartic terms in the electric superpotential,
and with $h \mu_i^2$ the coefficients of the linear 
deformations, corresponding
to the masses of the quarks in the electric description.
In (\ref{magnspot}) we wrote a single coupling $h m$, for all the different
mesons, considering all their masses of the same order.

The $b$ coefficients of the beta functions before dualization
are
\be
b_{i}=3N_i-N_{i-1}-N_{i+1} \qquad i=1,\dots,n
\ee
where $N_{0}=N_{r+1}=0$.
After the dualization 
the coefficients $\widetilde b$ for the beta functions in the internal nodes 
result
\begin{eqnarray} 
\label{btilde0}
\widetilde b_{2k} &=& 2 N_{2k+1} + 2 N_{2k-1} - 3 N_{2k}  \\
\label{btilde}
\widetilde b_{2k+1} &=& N_{2k}+N_{2k+2}-N_{2k+1}-2N_{2k-1}-2N_{2k+3} 
\end{eqnarray}
where k runs from $1$ to $\frac{n-1}{2}$, and
$N_{n+1}=N_{n+2}=0$.
For the external nodes we have
\be
\widetilde b_1=N_1+N_2-2 N_3 \qquad \widetilde b_n=N_n+N_{n-1}-2 N_{n-2} 
\ee

To visualize the resulting magnetic theory  
(\ref{magnspot})
we exhibit below the content of the magnetic dual theory
for an $A_5$ quiver, which encodes the relevant
features.
\begin{center}
\begin{tabular}{c}
    \includegraphics[width=13cm]{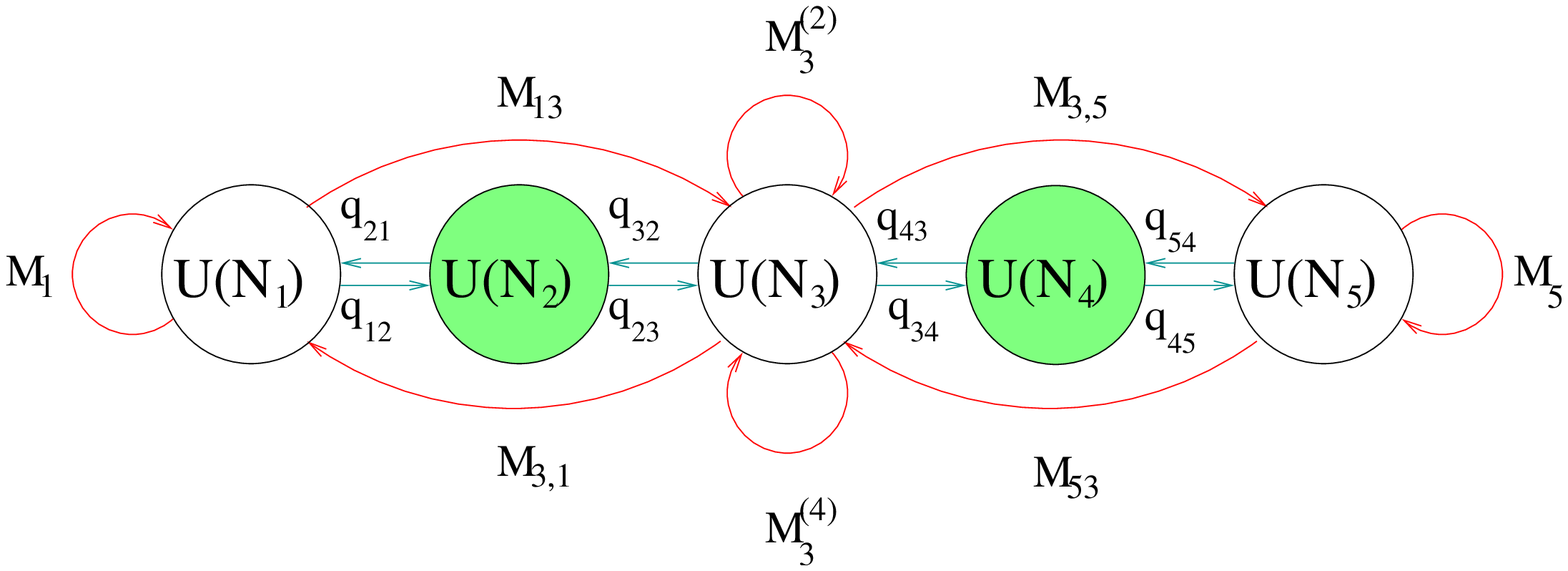}\\
\end{tabular}
\end{center}
The superpotential is
\begin{eqnarray}
W&=& h \left( M_{11} q_{12} q_{21}+M_{13} q_{32} q_{21}+M_{31}q_{12}q_{23}
+M_{33}^{(2)}q_{32}q_{23} \right) +
\nonumber\\ 
 &+& h 
\left( M_{33}^{(4)} q_{34} q_{43}+M_{35} q_{54} q_{43}+M_{53}q_{34}q_{45}
+M_{55}q_{54} q_{45}
\right )+
\nonumber \\
&+& h m \left( M_{11}^2+ M_{13}M_{31}+ {M_{33}^{(2)}}^2+
 M_{33}^{(2)}M_{33}^{(4)} 
+ {M_{33}^{(4)}}^2 +  M_{35}M_{53} + M_{55}^2  \right)+
\nonumber \\
&+&h  \left(\mu_1^2 M_{11}+\mu_{3,(2)}^2 M_{33}^{(2)}+
\mu_{3,(4)}^2 M_{33}^{(4)}+\mu_5^2 M_{55} \right)
\end{eqnarray}

\section{Metastable vacua in $A_3$ quivers}\label{metA3}
We start studying the existence and the slow decay
of non supersymmetric meta-stable vacua in $A_3$
quiver gauge theory, the simplest example
of an $A_n$ theory.
The $A_3$
gauge group is
$U(N_1)\times U(N_2)\times U(N_3)$. 
As already mentioned in section \ref{section2} for a $A_n$ theory, we 
 integrate out the adjoint fields and we perform
Seiberg duality on the central node under the constraint
\be
\label{SEIBA3}
N_2+1 \leq N_1+N_3< \frac{3}{2} N_2
\ee
The superpotential reads
\begin{eqnarray}\label{A3POT}
W &=& h \left(M_{1,1}q_{1,2}q_{2,1}+ M_{1,3}q_{3,2}q_{2,1}
+ M_{3,1}q_{1,2}q_{2,3} + M_{3,3}q_{3,2}q_{2,3}\right)+\nonumber \\
&+&h \mu_1^2 M_{1,1}+h \mu_3^2 M_{3,3}
\end{eqnarray}
where all the mass terms for the mesons have been neglected.
Turning on
these terms does not ruin the metastability
analysis at least for very small masses compared to
the supersymmetry breaking scale. 
Such deformations slightly shift the value of the pseudomoduli in
the non supersymmetric minimum, breaking R-symmetry \cite{Murayama1}.
We neglect them in the following.

The central node yields the magnetic gauge group
$U(N_1+N_3-N_2)$
 whereas the groups at the two external nodes 
are considered as flavour groups, much less coupled.
We discuss in section \ref{BETAF}
the consistency of this assumption.
Since the gauge group is IR free in the low energy description,
and the flavours are less coupled,
we are allowed to neglect Kahler corrections
and take it as canonical \cite{ISS}.
Moreover the $D$-term corrections to the one loop effective potential
due to the flavour nodes are negligible
with respect to the $F$-term corrections.

Now, there are two different choices of ranks for 
the $A_3$ theories, which can
give meta-stable vacua: the first possibility is that $N_1 < N_2 \leq  N_3$,
the second one is $N_1<N_2>N_3$.
We study separately the two cases which
show meta-stable vacua in a similar manner.

\subsection*{$N_1<N_2\leq N_3$}

We analyze here the case $N_1<N_2<N_3$; the equal ranks limit 
can be easily included. After the dualization
the ranks obey the following inequalities
$N_1<\widetilde N_2=N_1+N_3-N_2<N_3$.

We work in the regime where $|\mu_1|>|\mu_3|$,
and we comment on what happens in the opposite limit in the
appendix \ref{goldStone}, 
where we shall discuss dangerous tachyonic directions
in the quark fields.

We find that the following vacuum is a 
non supersymmetric 
tree level minimum
\begin{eqnarray} \label{vevs}
q_{1,2} = q_{2,1} &=& \mu_1 \left( \mathbf{1}_{N_1} \,\,\,\,\,\, 0 \right)
\,\,\,\,\,\,\,\,
q_{2,3}=q_{3,2}=\left(
    \begin{array}{cc}
    0&\mu_3 \mathbf 1_{\widetilde N_2-N_1}\\
    0&0
    \end{array} \right)
    \nonumber \\
M_{1,1}&=&0\,\,\,\,\,\,\,\, M_{1,3}= M_{3,1}=0 \,\,\,\,\,\,\, M_{3,3}=
\left(
    \begin{array}{cc}
    0&0\\
    0&X
    \end{array} \right)
\end{eqnarray}
where the field $X$ is the pseudomodulus,
which is a massless field not associated with any 
broken global symmetries.
This flat direction has to be stabilized by
the one loop corrections.
Westart the one loop analysis 
by rearranging the fields and expanding 
around the
vevs
\begin{eqnarray}
q = \left(
\begin{array}{c}
q_{1,2}\\
\hline
q_{3,2}\\
\end{array}
\right)&=& \left(
\begin{array}{cc}
\mu_1+\Sigma_1&\Sigma2\\
\hline
\Sigma_3&\mu_3+\Sigma_4\\
\Phi_1&\Phi_2\\
\end{array}
\right) \,\,\,\,\,\,\, \widetilde q = \left(
\begin{array}{c|c}
q_{2,1}&q_{2,3}\\
\end{array}
\right)= \left(
\begin{array}{c|cc}
\mu_1+\Sigma_5&\Sigma_6&\Phi_3 \\
\Sigma_7&\mu_3+\Sigma_8&\Phi_4 \\
\end{array}
\right)\nonumber \\
M &=& \left(
\begin{array}{c|c}
M_{1,1}&M_{1,3}\\
\hline
M_{3,1}&M_{3,3}\\
\end{array}
\right)= \left(
\begin{array}{c|cc}
\Sigma_{9}&\Sigma_{10}&\Phi_5\\
\hline
\Sigma_{11}&\Sigma_{13}&\Phi_6 \\
\Phi_7&\Phi_8&X+\Sigma\\
\end{array}
\right)
\end{eqnarray}
We now compute the superpotential 
at the second order in the fluctuations. We
find that the non supersymmetric sector is a set of decoupled
O'Raifeartaigh like models
with superpotential
\begin{equation}
W = h \mu_3^2 X + h X (\Phi_1 \Phi_3 + \Phi_2 \Phi_4) +h \mu_3 (\Phi_1
\Phi_5 +\Phi_2 \Phi_6) + h \mu_1(\Phi_3 \Phi_7 +\Phi_4 \Phi_8)
\end{equation}
In this way all the pseudomoduli can get a mass.
The quantum corrections behave
exactly as in \cite{ISS}, which means that the pseudomoduli get
positive squared mass around the origin of the field space. 

The choice (\ref{vevs}) guarantees that there are no tachyonic directions
and have to be made coherently with the hierarchy of the couplings
$\mu_i$; see the Appendix \ref{goldStone} for details.\\

The lifetime of the non supersymmetric vacuum is related to
the value of the scalar potential in the minimum, and to the
displacement of the vevs of the fields between the 
false and the true vacuum.
The scalar potential in the non supersymmetric minimum is
\be
\label{scpot1}
V_{min} = (N_3+N_1-\widetilde N_2) |h \mu_3^2|^2 = N_2 |h \mu_3^2|^2
\ee

The vevs of the fields in the supersymmetric vacuum have to be
studied considering 
the non perturbative contributions arising from gaugino
condensation.
When we
take into account these non perturbative effects, we expect that the
mesons get large vevs
and this allows
us to integrate out the quarks using their equation of motion,
$q_{i,j}=0$. In the supersymmetric vacua also $M_{1,3}=0$ and
$M_{3,1}=0$. If we define
\begin{equation} \label{bigmes}
M = \left(
\begin{array}{cc}
M_{1,1} & 0 \\
0 & M_{3,3}\\
\end{array}
\right)
\end{equation}
the effective superpotential is
\begin{equation} \label{W_dyn}
W = (N_1+N_3-N_2)\left(\det(h M) \Lambda_{2i}^{2 N_1+2 N_3-3 N_2}
\right)^{\frac{1}{N_1+N_3-N_2}} -h\left( \mu_1^2 tr{M_{1,1}}+\mu_3^2 
tr{M_{3,3}} \right)
\end{equation}
We have now to solve the equation of motion for $M_1$ and $M_3$.
The equations
to be solved are
\begin{eqnarray} \label{eq_M1-M_3}
\left( h^M M_{1,1}^{(N_2-N_3)}M_{3,3}^{N_3}
\Lambda_{2i}^{(2 N_1+2 N_3-3 N_2)}\right)^{\frac{1}{N_1+N_3-N_2}}
- \mu_1^2 &=& 0\nonumber\\
\left( h^{N_2} M_{1,1}^{N_1}M_{3,3}^{(N_2-N_1)}
\Lambda_{2i}^{(2 N_1+2 N_3-3 N_2)}
\right)^
{\frac{1}{N_1+N_3-N_2}} - \mu_3^2 &=& 0
\end{eqnarray}
The vevs of the mesons follow solving (\ref{eq_M1-M_3})
\begin{equation} \label{susyvev}
\langle h M_{1,1} 
\rangle = \mu_1^{2\frac{N_1-N_2}{N_2}} \mu_3^{2\frac{N_3}{N_2}}
\Lambda_{2i}^{\frac{3 N_2-2 N_3-2 N_1}{N_2}} \mathbf 1_{N_1}\,\,\,\,\,\,\,
\,\,\,
\langle h M_{3,3} 
\rangle = \mu_1^{2\frac{N_1}{N_2}} \mu_3^{2\frac{N_3-N_2}{N_2}}
\Lambda_{2i}^{\frac{3 N_2-2 N_3-2 N_1}{N_2}} \mathbf 1_{N_3}
\end{equation}
Since $|\mu_1|>|\mu_3|$, it follows that 
$\langle h M_{3,3} \rangle > \langle h M_{1,1} \rangle$.
This implies that in the evaluation of the bounce action, with the 
triangular barrier 
\cite{Duncan}, 
we can consider only the displacement of $M_3$
in the field space. We obtain for the bounce action
\begin{equation}
S \sim \frac{(\Delta \Phi)^4}{\Delta V}=
\left(\frac{\mu_1}{\mu_3}\right)^{\frac{3 N_2-2 N_3}{N_2}}
\left(\frac{\Lambda_{2i}}{\mu_1}\right)^{4\frac{3 N_2-2 N_3-2 N_1}{N_2}} 
\end{equation}
Both exponents are positive in the range (\ref{SEIBA3}).
This implies that $S_B \gg 1$, and the vacuum is long living.

\subsection*{$N_1<N_2>N_3$}
The ranks of the groups after the duality obey the relation 
$N_1>\widetilde N_2=N_1+N_3-N_2 < N_3$.
We choose now $|\mu_1|>|\mu_3|$, 
but we show in the appendix \ref{goldStone} that 
also the other choice is possible, leading to other vacua. 
In the meta-stable vacuum all the vevs of the fields have to be chosen to be 
zero except a block of the quarks $q_{1,2}$ and $q_{2,1}$ and the 
pseudomoduli. The vevs are
\begin{equation}
\label{vacuum2}
q_{1,2}= \mu_1 \left(
\begin{array}{c}
\mathbf 1_{N_1}\\
\mathbf 0\\
\end{array}
\right) \,\,\,\,\,\,\,\,\,\,\,\, q_{2,1}^T=\mu_1 \left(
\begin{array}{c}
\mathbf 1_{N_1}\\
\mathbf 0\\
\end{array}
\right)
\end{equation}
The pseudomoduli come out from the meson $M_{3,3}$ and a $(\widetilde
N_2-N_1)\times(\widetilde N_2 -N_1)$ 
diagonal block of the other meson, $M_{1,1}$.
The one loop analysis is the same as before and 
lifts all the flat directions.

In order to estimate the lifetime we 
need the vevs of the fields in the supersymmetric vacuum, which
are again (\ref{susyvev}), and the value of the scalar potential
in the non supersymmetric vacuum (\ref{vacuum2})
\be
V_{min}=(N_2-N_3)|h \mu_1|^2+N_3 |h \mu_3|^2
\ee
Since $|\mu_1|>|\mu_3|$
we approximate the scalar potential by the term $\sim |\mu_1|^2$
and the field displacement by $\langle h M_3 \rangle$,
obtaining as bounce action
\be
S \sim \left(\frac{\mu_1}{\mu_3} \right)^{2 \frac{N_2-N_3}{N_2}}
\left (\frac{\Lambda_{2i}}{\mu_1} \right)^{4\frac{3 N_2-2 N_1-2 N_3}{N_2}}
\gg 1
\ee

\section{Renormalization group flow}\label{BETAF}
The analysis of 
sections \ref{alternate} and \ref{metA3}
relies on the fact that we neglect 
the contributions to the dynamics due to the odd nodes.
It means that these groups have to be treated as 
flavours groups, i.e. global symmetries.
However, in the $A_n$ quiver theory 
each node represents a gauge group factor and we have to
analyze how its coupling runs with the energy.

The magnetic window (\ref{magnwind}) constraints the even nodes
to be UV free in the high energy description, i.e. $b_{2i}>0$.
The odd groups are not uniquely determined by (\ref{magnwind})
and can be both UV free or IR free in the electric description.
In the first case we will choose their scale $\Lambda_{2i+1}$
to be much lower than the even one 
\be
\label{SCALE1}
\Lambda_{2i+1}\ll \Lambda_{2i}.
\ee
In the
second case, when $b_{2i+1}<0$, $\Lambda_{2i+1}$ is a Landau pole
and we take
\be
\label{SCALE2}
\Lambda_{2i+1}\gg \Lambda_{2i}.
\ee
In these regimes 
the even nodes become strongly coupled before 
the odd ones in the flow toward the infrared.
This means that we need a new description
provided by Seiberg dualities on the even nodes.

In order to trust the perturbative description
at low energy,
we have to impose that at the supersymmetry breaking scale 
(typically $\mu_i$)
the odd nodes (flavour),
are less coupled than the even ones (gauge),
which are always IR free.
This requirement will give other constraints on
the scales.

As already said there are two possible behaviors
of the flavour groups above the scale $\Lambda_{2i}$:
they can be IR free or UV free.
For both cases 
there are three different possibilities
about the beta coefficients in the low energy description.

We start discussing the case when the flavours group are UV free
in the electric description.  
The following 
three possibilities arise for each flavour group 
$U(N_{2k+1})$ in the dual theory (Plots 1,2,3 in Figure 1).

\begin{enumerate}
\item The first one
is characterized by
\be
b_{2k+1}>0 \qquad  \qquad \widetilde b_{2k+1}<\widetilde b_{2i}<0 
\label{IRfre}
\ee
In this case the flavour groups $U(N_{2k+1})$
are more IR free than the even nodes 
after Seiberg duality.
The couplings of the flavour groups
become more and more smaller than the
couplings of the gauge groups along the flow
toward low energy.
Hence we do not need other constraints on the scales
except (\ref{SCALE1}).
\item 
The second possibility is reported in Plot 2 in Figure 1 
\be 
\label{seccase}
b_{2k+1}>0 \qquad  \qquad
\widetilde b_{2i}<\widetilde b_{2k+1}<0
\ee
The flavour groups $U(N_{2k+1})$ are IR free
in the dual theory,
but less than the $U(\widetilde N_{2i})$ gauge groups (\ref{seccase}).
Below a certain energy scale the flavours become
more coupled than the gauge groups. If this happens
before the supersymmetry breaking scale we cannot
trust our description anymore.
To solve this problem we have to choose the correct hierarchy
between the electric scales of the flavour and the gauge groups,
and the supersymmetry breaking scale.
We impose that the couplings of the flavours are smaller
than the couplings of the gauge groups at the breaking scale,
in the magnetic description.
This condition can be rewritten in terms
of electric scales only using
the matching between the magnetic and the electric scales
of the flavours. This procedure is explained
in the Appendix B and gives the following condition on 
$\Lambda_{2k+1}$
\be
\label{condscales}
\Lambda_{2k+1}\ll 
\left( \frac{\mu}{\Lambda_{2i}} \right)^{\frac{\widetilde b_{2k+1}-
\widetilde b_{2i}}{b_{2k+1}}}  \Lambda_{2i} 
\ll \Lambda_{2i}
\ee
This imposes a constraint stronger than (\ref{SCALE1}) on the 
strong coupling scale of the flavours.

\item The third possibility (Plot 3 Figure 1) is 
\be
b_{2k+1}>0 \qquad  \qquad 
\widetilde b_{2k+1}>0 \label{UVfre}
\ee
In this case the flavour group $U(N_{2k+1})$ is asymptotically
free in the low energy description. 
Once again we have to impose that at the breaking scale the
flavours are less coupled than the gauge groups. The procedure
is the same outlined above, and the condition is the same as 
(\ref{condscales}).
This case may become problematic in the far infrared. 
Indeed, since the flavour group is UV free, it develops
strong dynamics at low energy.
If we take into account the non perturbative contributions
they could restore supersymmetry.
Another interesting feature
is the appearance of cascading gauge theories, flowing
in the IR. 
We do not discuss these issues here.

\end{enumerate}

If the flavour groups $U(N_{2k+1})$ are IR free in the electric description
the same three possibilities 
discussed above arise (see Plots $4$, $5$, and $6$ of 
Figure 1).
\begin{enumerate}
\item[4.]
The plot $4$ of Figure 1 is characterized by
\be
b_{2k+1}<0 \qquad  \qquad 
\widetilde b_{2k+1}<\widetilde b_{2i}<0 
\ee 
Here we do not need any other constraint 
except (\ref{SCALE2}).

\item[5.]
The plot $5$ in Figure 1 is 
\be
b_{2k+1}<0 \qquad  \qquad 
\widetilde b_{2i}<\widetilde b_{2k+1}<0 
\ee 
The requirement that the odd nodes
are less coupled than the even ones at the 
supersymmetry breaking scale give once again
non trivial constraints, with the same procedure outlined previously
\be
\label{condscalebis}
\Lambda_{2k+1}\gg 
\left( \frac{\Lambda_{2i}}{\mu} \right)^{\frac{\widetilde b_{2i}-\widetilde 
b_{2k+1}}{b_{2k+1}}}  \Lambda_{2i} 
\gg \Lambda_{2i}
\ee
where now the strong coupling scale
of the flavour groups 
in the electric description is a Landau pole.

\item[6.] 
The last possibility (Plot 6 of Figure 1)
\be
b_{2k+1}<0 \qquad  \qquad 
\widetilde b_{2k+1}>0 
\ee 
lead to the same constraint (\ref{condscalebis}). 
In the far infrared  the strong dynamics of the flavours node
can lead to non perturbative phenomena, as in the case 3.

\end{enumerate}
 
\begin{center}
\begin{tabular}{c|c}
    \includegraphics[width=8cm]{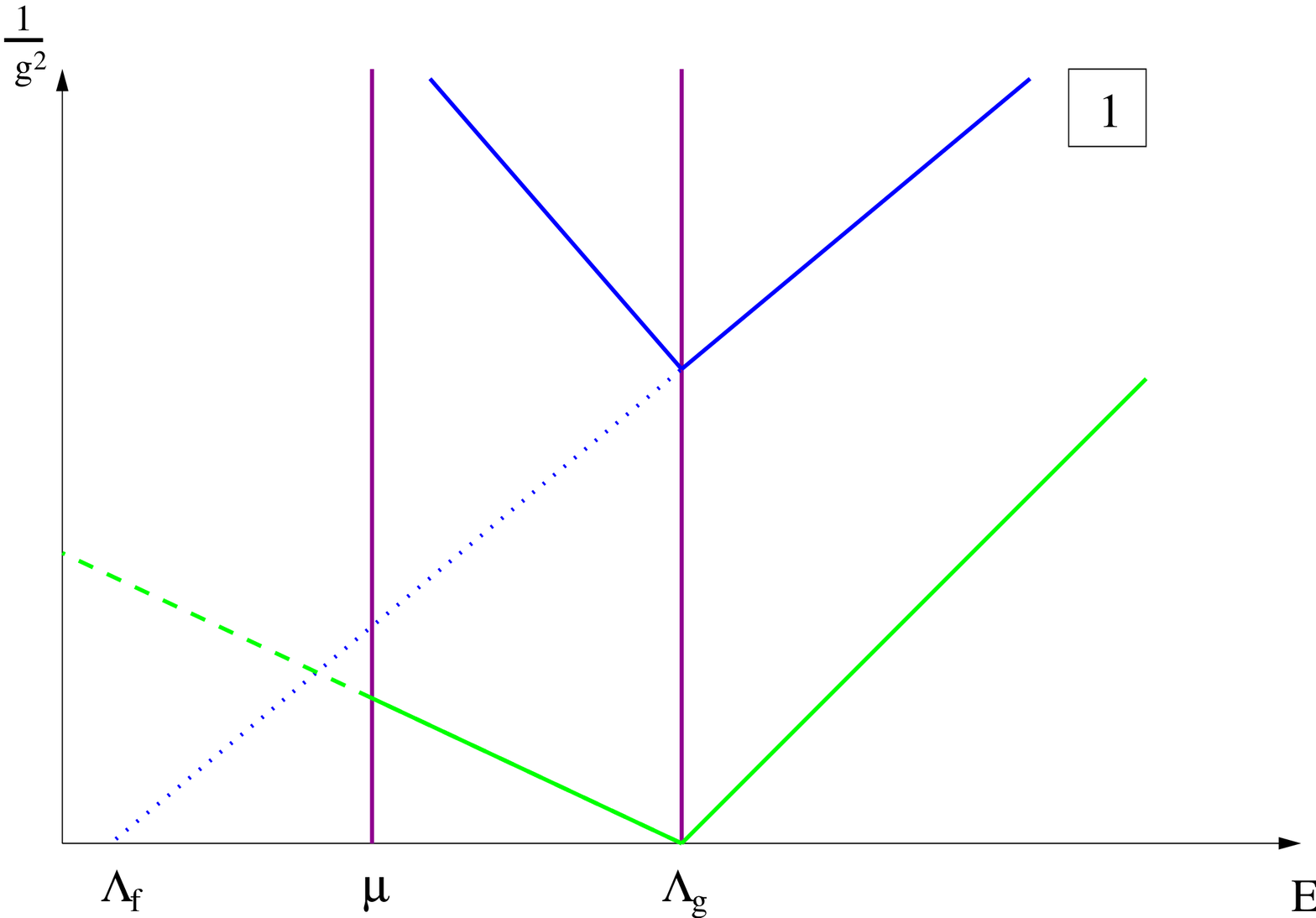}&
    \includegraphics[width=8cm]{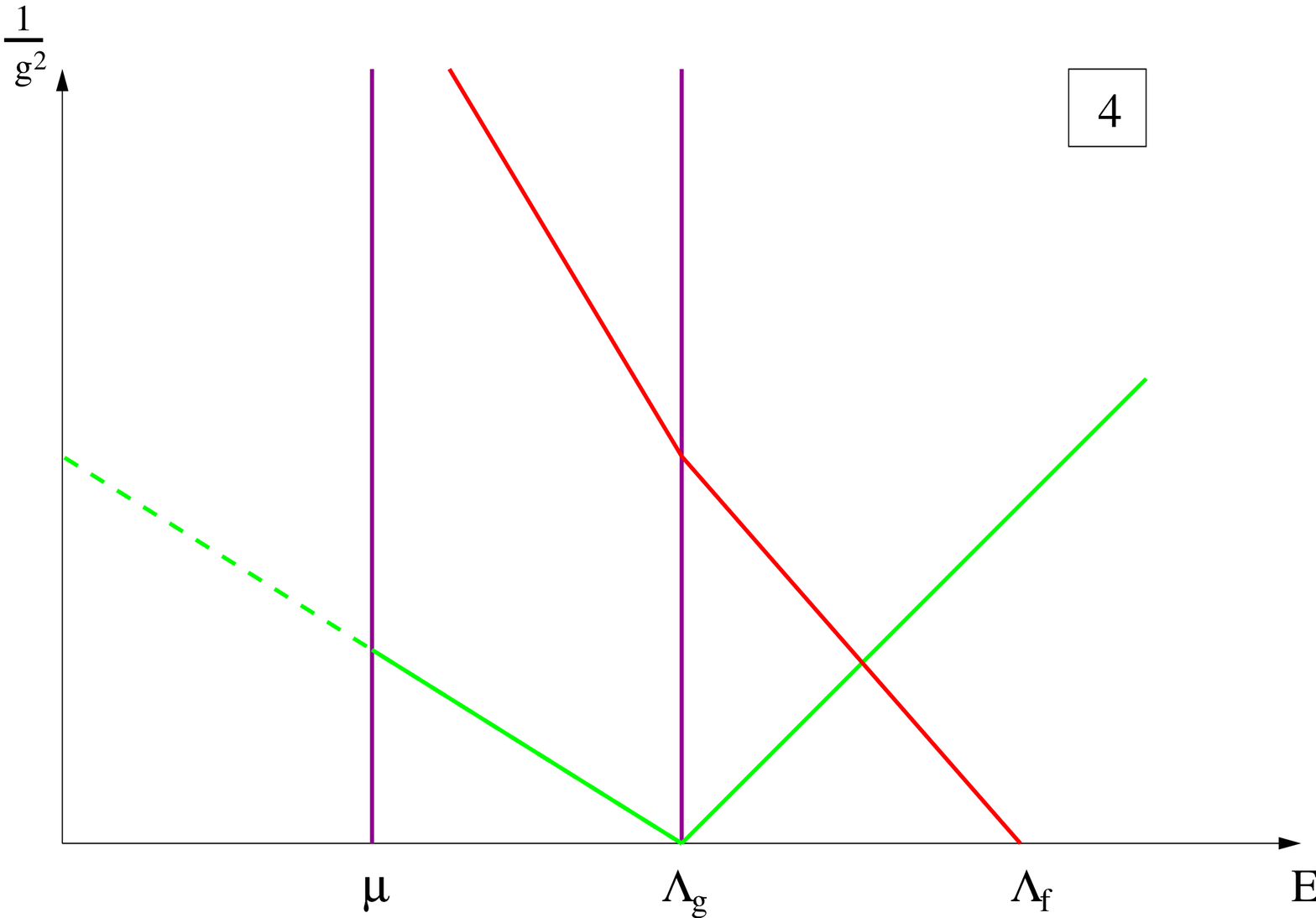}\\
\hline
   \includegraphics[width=8cm]{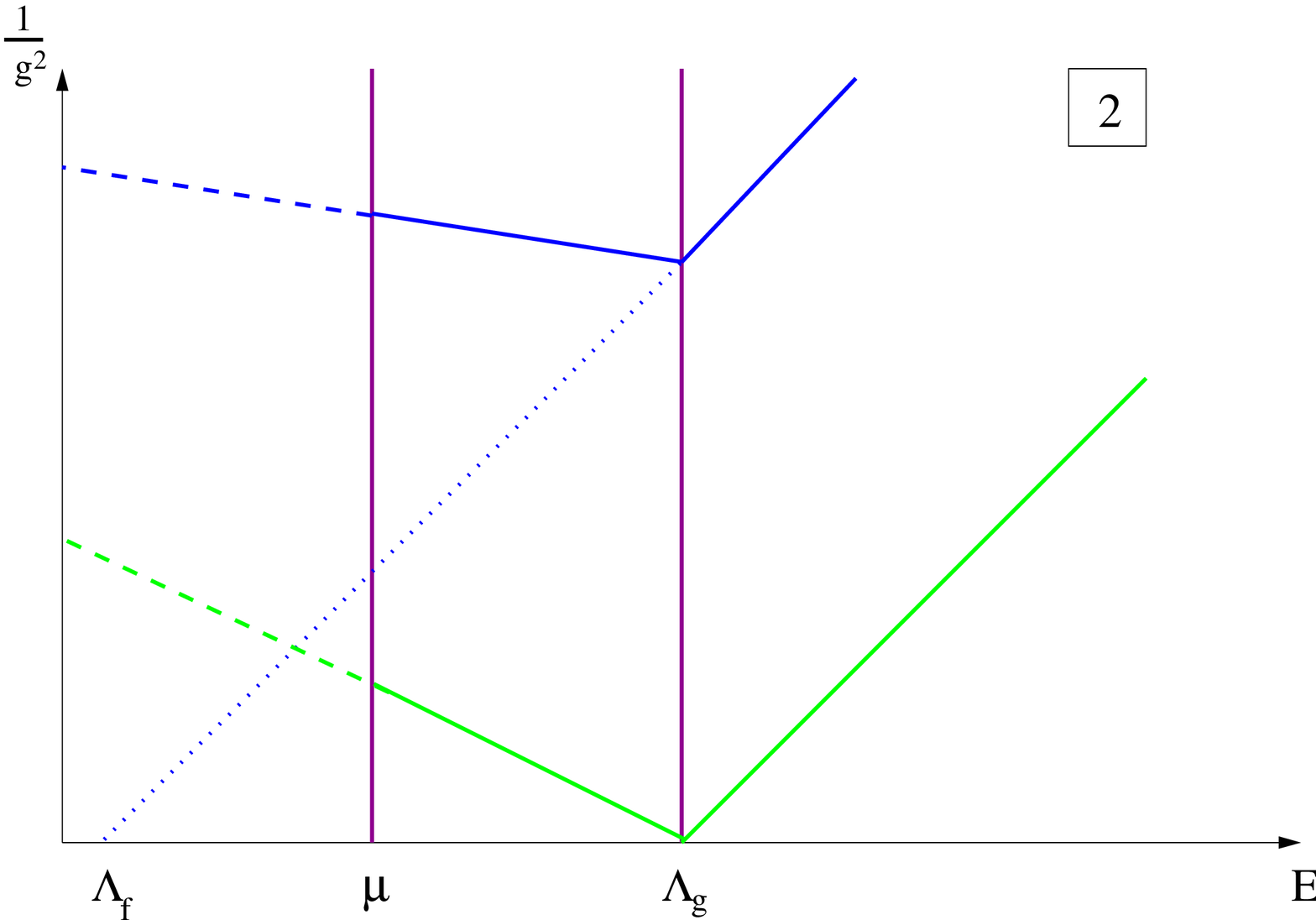}&
      \includegraphics[width=8cm]{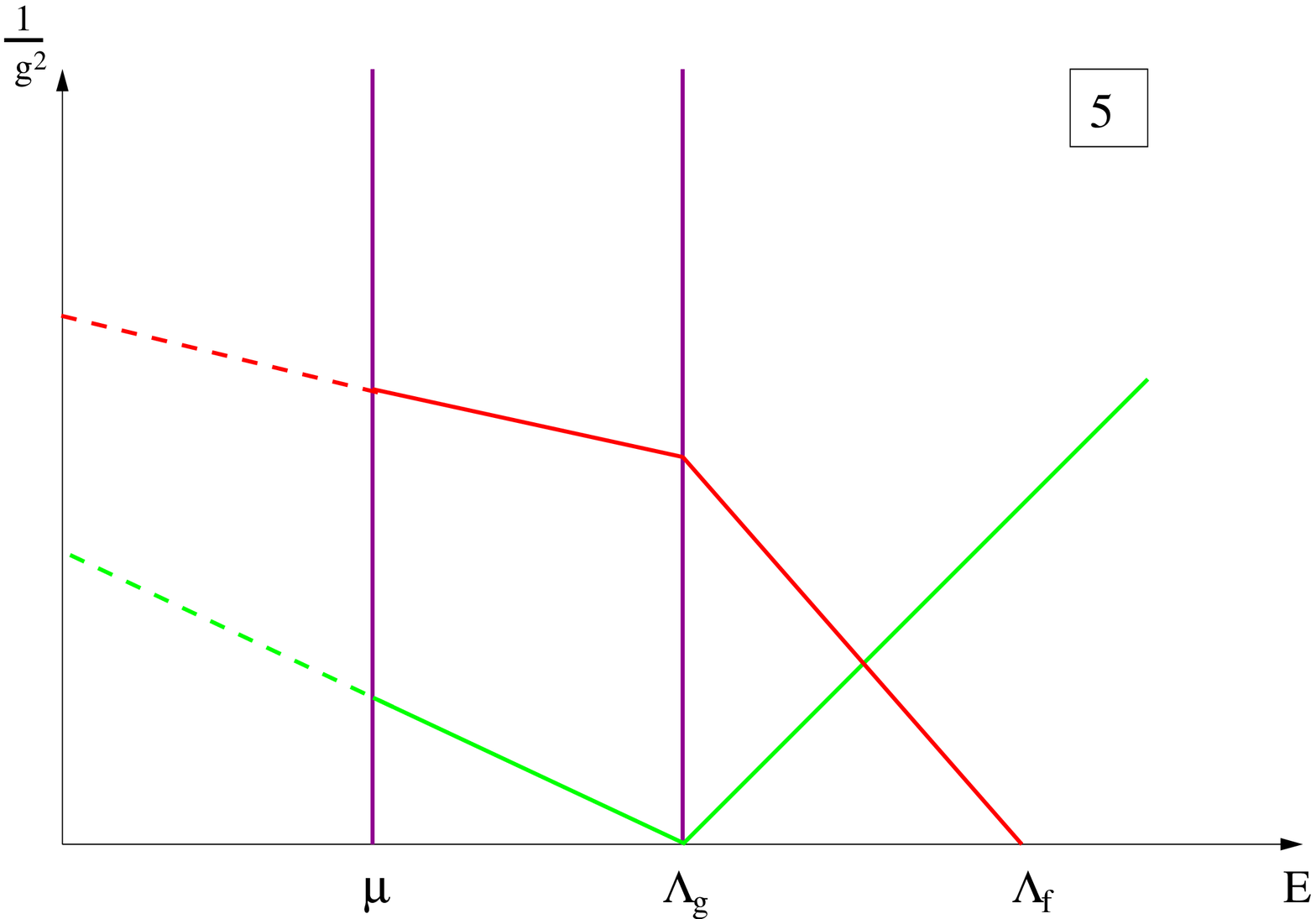}\\
\hline
   \includegraphics[width=8cm]{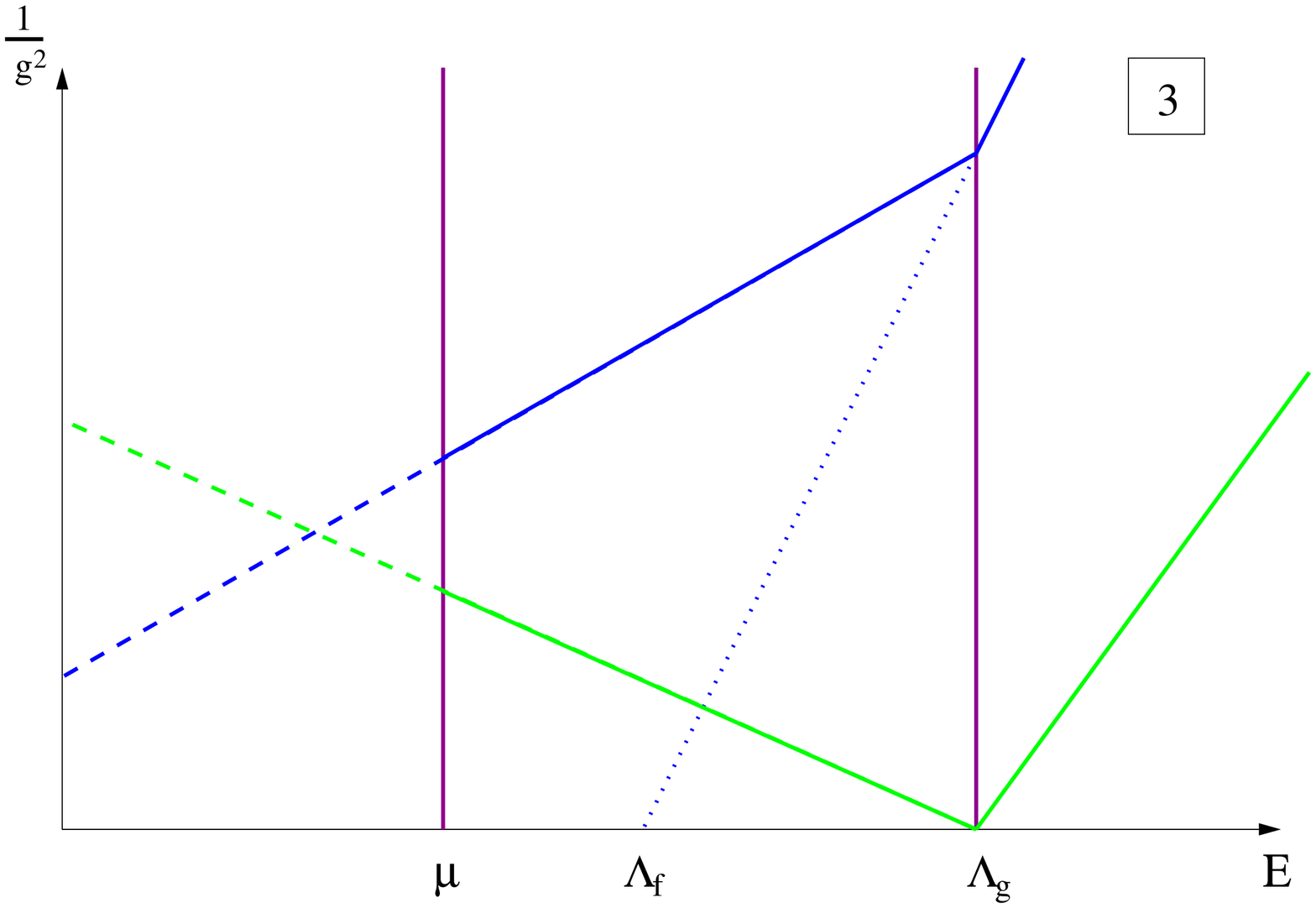}&
      \includegraphics[width=8cm]{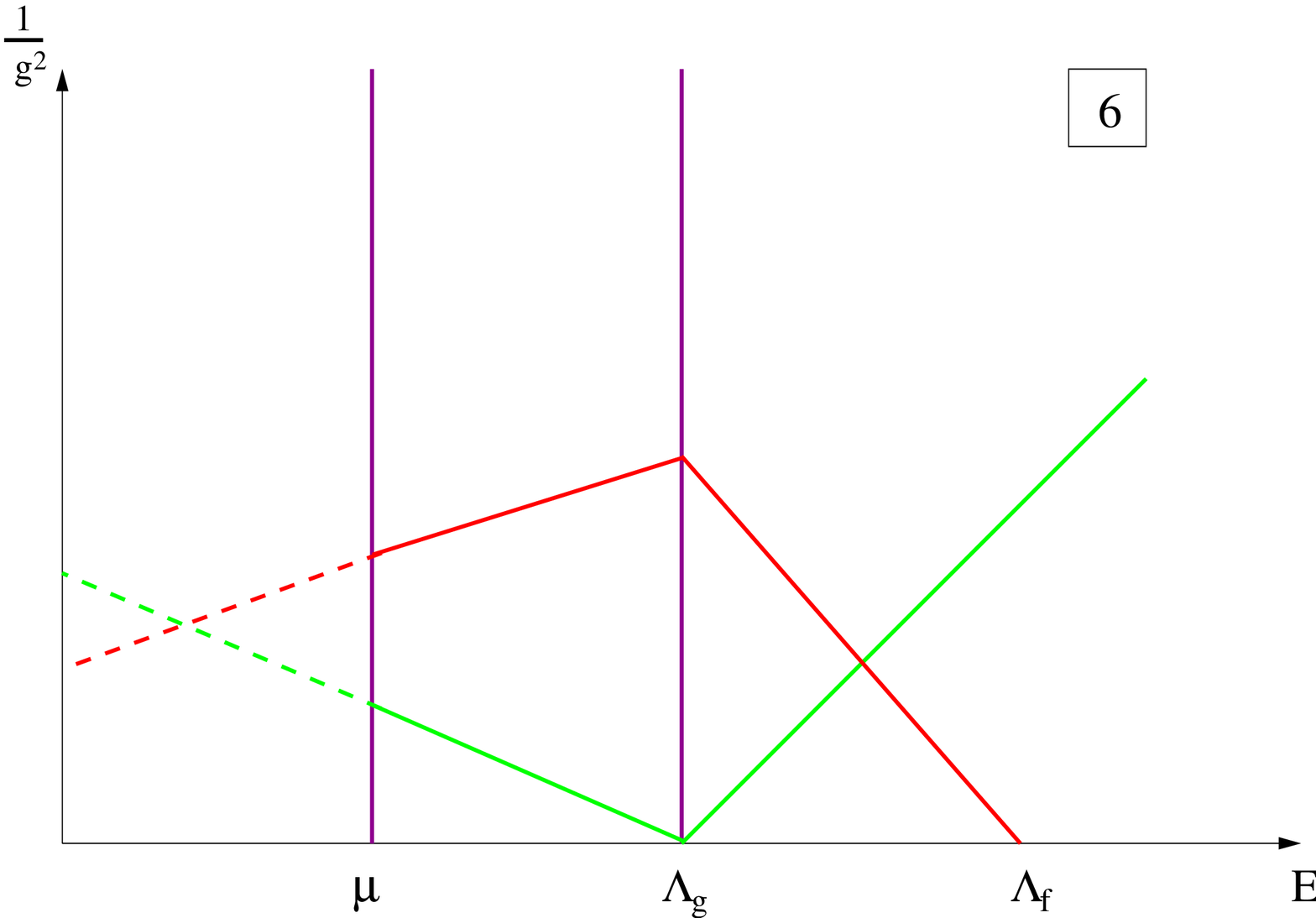}\\
\end{tabular}
\end{center}
Figure 1: \emph{The blue lines refer to flavour/odd 
groups which are UV free in the electric description,
while the red ones are IR free. 
The green lines refer to the gauge/even group couplings. We denote
with $\mu$ the supersymmetry breaking scale, and
$\Lambda_{G}$ and $\Lambda_{F}$ are the strong 
coupling scales of the gauge and the flavour groups,
respectively}.

\section{Meta-stable $A_n$} \label{extension}

We work in the regime where
the ratio $\frac{\mu_i^2}{m}$ is 
larger than the strong scale
of the even nodes $\Lambda_{2i}$.
This requirement is satisfied 
if $\lambda_{i} \gg \Lambda_{2i}^2$ in the electric theory.
This allows us to ignore in 
the dual superpotential (\ref{magnspot})
the presence of quadratic
deformations in the mesonic fields.

In this approximation the superpotential
of the $A_n$ quiver
(\ref{magnspot}) 
reduces to $\frac{n-1}{2}$ 
copies of $A_3$ superpotentials.
Hence a generic $A_n$ diagram results 
decomposable in copies of $A_3$
quivers, where every adjacent 
pair shares an odd node.

For each $A_3$ the even nodes provide 
the magnetic gauge groups,
and each $A_3$ has long living 
metastable vacua,
if the perturbative 
window is correct. 
It follows that the $A_n$ quiver theory, which is 
a set of metastable $A_3$ quivers,
possesses metastable vacua.

We still have to be sure 
of the perturbative regime.
This means that we have to 
control the gauge contributions from the
odd nodes of the $A_n$ diagram.
We have to proceed as in section \ref{BETAF},
and 
study the beta coefficients
of the groups.
From (\ref{btilde}) we can see that the magnetic beta coefficients
of the internal odd nodes involve 
the ranks of the next to next neighbor groups,
i.e. they depend on five integer numbers.
This means that in order to know these beta coefficients
it is enough to study the $A_5$ consistent with (\ref{magnwind}).
In the appendix \ref{acinque} we classify
all the possible metastable $A_5$ diagrams 
and we
give the corresponding electric and 
magnetic beta coefficients
of the central flavour node. 
This classification describes the RG behaviour of all the internal 
odd nodes of the $A_n$.

The running of the first
and of the $n$-th node of the $A_n$ quiver 
is still undefined and it is
discussed in the appendix \ref{acinque}.

This provides a classification of metastable
$A_n$ quiver gauge theories 
with alternate Seiberg dualities.

\subsection{Example }
We show now a simple example of metastable $A_n$ diagram.
We choose the even nodes in the electric description
to become strongly coupled at the same scale $\Lambda_{2i}$.
We require that at such scale the flavours (odd nodes) are
less coupled than the gauge ones.  
Moreover we will show that we can also 
require that in the low energy description
all the nodes are IR free 
and also that the flavour groups
(odd nodes) are less coupled than the gauge groups (even nodes)
at any scale below the $\Lambda_{2i}$.

We study an $A_n$ theory, where $n = 4
k +1$, with $k$ integer. The chain is built as follow
\begin{center}
\begin{tabular}{c}
    \includegraphics[width=15cm]{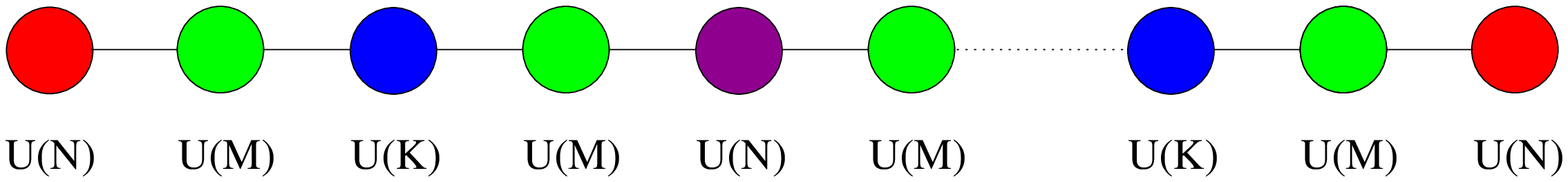}\\
\end{tabular}
\end{center}
with $N<M<K$. This range allows for metastable vacuum in each
$A_3$ piece as showed previously.
We perform alternate Seiberg dualities,
working in the 
in the window
\be \label{dis1}
  M +1 < N+K < \frac{3}{2}M \nonumber 
\ee

Thanks to the simple choice for the ranks we
have four values for the $b$ coefficients of the beta functions in
the electric description, and four values for the coefficients
$\widetilde b$. They are summarized in the following table
\begin{center} \label{tab1}
\begin{tabular}{c|c|cc}
   node& $b$ &$\widetilde b$ & \\
  \hline
  $1,n$ (red)  & $3N-M$  & $N-2K+M$&  \\
  \hline
  $2i$ (green)  & $3M-N-K$& $2K+2N-3M$ & \\
  \hline
  $4i-1$ (blue) & $3K-2M$ & $2M-4N-K$ &  $i =1,\dots, \frac{n-1}{4}$\\
  \hline
  $4j+1$ (violet) & $3N-2M$ & $2M-4K-N$ & $j =1,\dots, \frac{n-5}{4}$  \\
\end{tabular}
\end{center}

We require that in the magnetic description 
all the nodes are
IR free. Moreover we require the beta coefficients 
of the odd groups 
to be lower than the even group ones,
i.e. $\widetilde b_{\text{odd}}<\widetilde b_{2i}$. 
This restricts the window to
\be
K> 2N \qquad  \qquad 3N < 2M< 4N+K
\ee
In this regime all the nodes in the electric description are
UV free except the $4j+1$-th ones.
Seiberg duality is allowed on the even nodes,
if we impose the following hierarchy of scales 
\begin{equation}\label{gerarchia}
    \Lambda_{1},\Lambda_{n},\Lambda_{4i-1} \ll  \Lambda_{2i} \ll \Lambda_{4j+1}
\end{equation}
The running of the gauge couplings of the different nodes
are depicted in Figure 2.
\begin{center}
\begin{tabular}{c}
   \includegraphics[width=6cm]{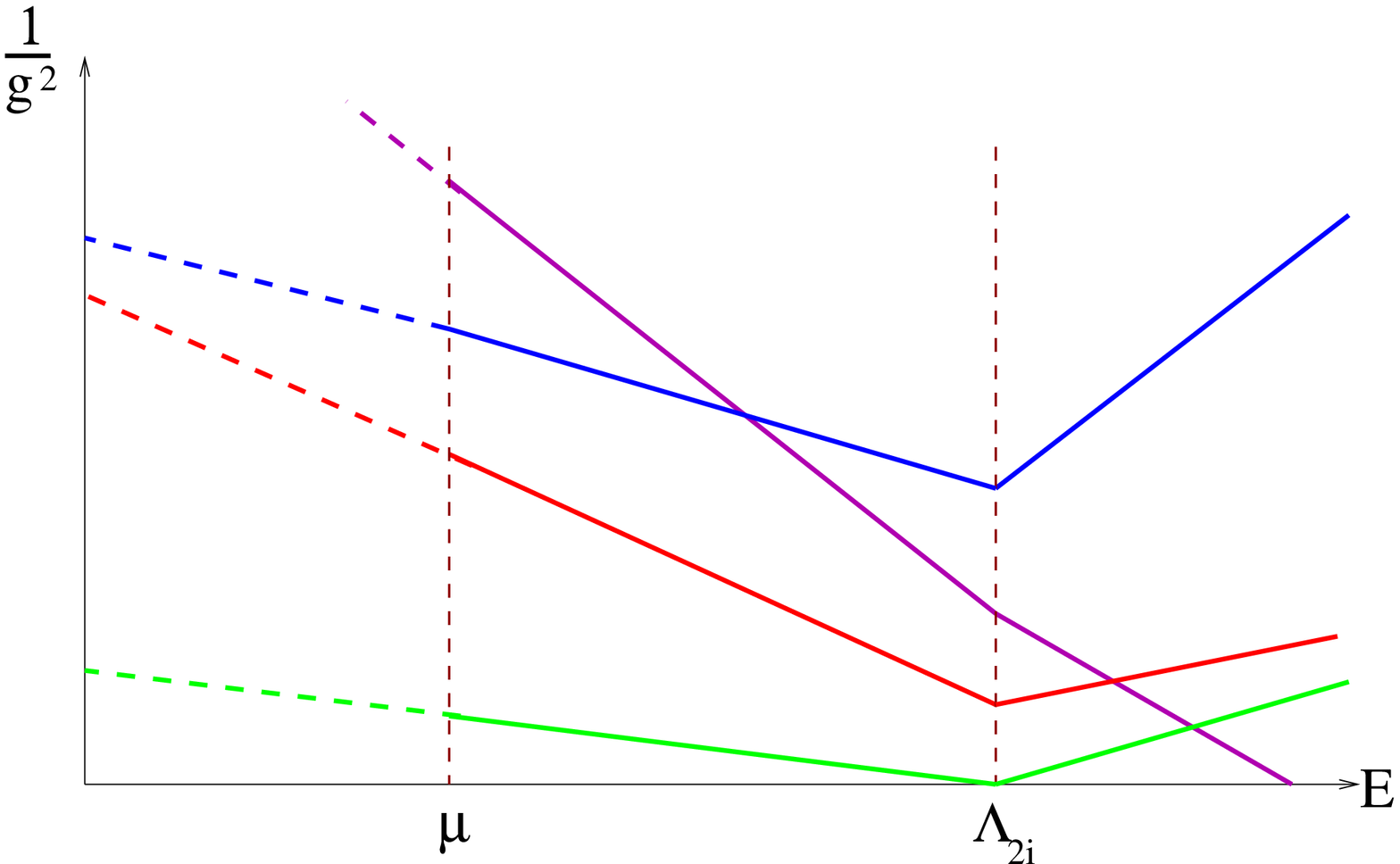}
\end{tabular}
\end{center}
Figure 2: \emph{The green line represents the running of the coupling of
the even sites. The violet line is related to the $4j+1$-th sites,
the blue one to the $4i-1$-th sites and the red to the first and
the last nodes.}
\\
\\
At high energy the $4j+1$-th nodes are
strongly coupled,
while the other nodes are all UV free. 
At the scale $\Lambda_{2i}$ the even nodes become
strongly coupled and Seiberg dualities take place.
All the runnings of the couplings are changed by these dualities,
and all the coefficients of the beta functions $\widetilde b_i$ become negative.
Hence at energy scale lower than $\Lambda_{2i}$ the
theory is weakly coupled. Furthermore 
the beta coefficients of the odd nodes
are more negative than the even node ones. 
This guarantees that we can rely on perturbative computations,
treating the odd nodes as flavours.

\section{Gauge mediation}

The models analyzed in this work can 
admit mechanisms of gauge mediation.
This means that the breaking of supersymmetry can be transmitted to 
the Standard Model sector via a gauge interaction.
This idea has already appeared in the literature of metastable vacua 
in $A_n$ theories \cite{Kitano,Kawano}.

Different realizations are possible here. A first one, of
direct gauge mediation, identifies the SM gauge group with a subgroup
of a flavour group in the quiver \cite{Kitano} and leads to a gaugino
mass consistently with the bound of \cite{Murayama1}.

A second possibility \cite{Kawano} is to connect one of the extremal nodes of 
the $A_n$ quiver with a new gauge group, which represents
the Standard Model gauge group.
The arrows connecting these nodes are associated with the messengers
$f$ and $\widetilde f$, which communicate the breaking of supersymmetry to the
standard model.  
Neglecting all the
quartic terms, except the term which couples the messengers $f ,\widetilde
f$ with the last meson, it is possible to show that also in this
case gaugino masses arise at one loop.

In our models of metastable $A_n$ quivers   
another possibility arises for gauge mediation. 
It consists in substituting
an even node with the Standard Model gauge group.
\begin{center}
\begin{tabular}{c}
\includegraphics[width=15cm]{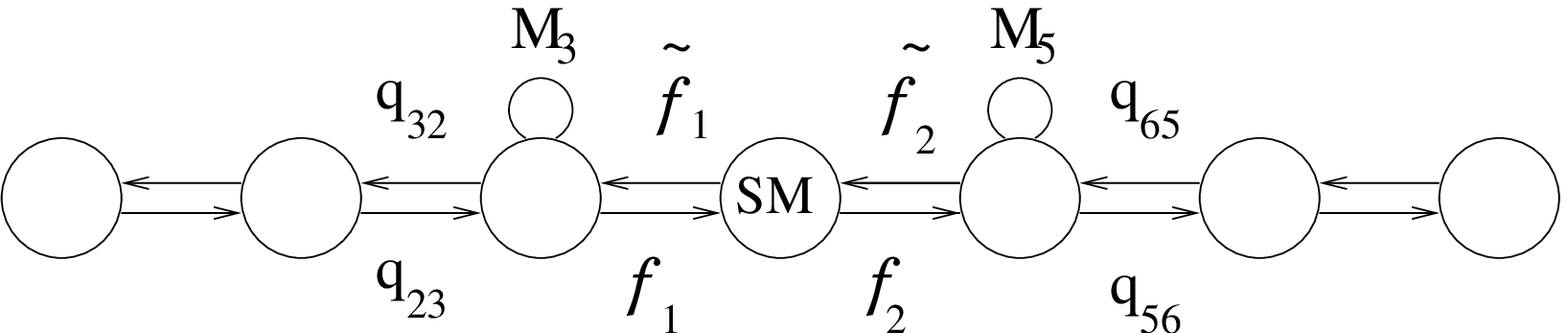}\\
\end{tabular}
\end{center}
The low energy
description is constituted by two metastable $A_n$ 
($A_3$ in this case)  
which are connected through the SM sector. 
Both communicate the supersymmetry breaking to the standard model.
The superpotential leads to two copies of messengers fields
related to the two different hidden sectors
\be
W=\left(m_1 + \theta^2 h_1 F_{M_3}  \right) f_1  \widetilde f_1+
\left(m_2 + \theta^2 h_2  F_{M_5} \right) f_2  \widetilde f_2
\ee
A gaugino mass arises at one loop proportional to
$\left(h_1 \frac{F_{M_3}}{m_1}+ h_2 \frac{F_{M_5}}{m_2}\right)$.

\section*{Conclusions}
We have studied metastability in models of $A_n$ quiver gauge
theories.  The low energy description in terms of macroscopic fields
can be achieved via Seiberg dualities at chosen nodes in the $A_n$
diagram. This choice defines, to a certain extent, the models.

A strategy for building acceptable models unfolds from the request for 
a reliable perturbative analysis.
This constrains the ranks of the gauge groups associated with the nodes
and their strong coupling scales.
We chose to dualize alternate nodes and we fixed two scales: a unique
breaking scale $\mu$ and a common strong coupling scale
$\Lambda_G$ for each dualized node.
The RG flows of the dualized and non dualized gauge groups must be
such that at  energy scale higher than
$\mu$ the gauge groups of the dualized nodes are more 
coupled than the other ones.

The RG properties of the different nodes of an $A_n$ 
quiver can be studied decomposing it in $A_5$ quivers
and the decomposition of the $A_n$ in $A_3$ patches 
gives the structure of the metastable vacuum.
In this way we classify all the possible 
$A_n$ quiver gauge theories which show
metastable vacua 
with the technique of alternating Seiberg dualities.

Finally we have discussed different patterns of gauge mediation.

\section*{Acknowledgments}
We would like to thank A.Butti, D.Forcella and A. Zaffaroni for
comments. We thank the GGI Center of Physics in Florence
where part of this work was done. 
This work has been supported in part by INFN, by PRIN prot.2005024045-002 
and the European Commission RTN program MRTN-CT-2004-005104.

\appendix

\section{Goldstone bosons}\label{goldStone}
The analysis we made in the $A_3$ theories started from the limit
$|\mu_1|>|\mu_3|$. Also the opposite limit can 
give 
meta-stable vacua. 
To understand the differences among the various choices,
we have to study the classical
masses acquired by the fields  
expanding them
around their vevs.

We study the case with ranks $N_1<\widetilde N_2<N_3$.
Since the flavor symmetry is $U(N_1)\times U(N_3)$, and not $U(N_1+N_3)$, the
linear terms of the mesons are different. 
We are still free to choose the hierarchy between them. 
We here analyze the breaking of the global symmetries
taking $|\mu_1| > |\mu_3|$. Treating
the gauge symmetry as a global one, and rearranging the quarks in
the form
\begin{equation}\label{quarks}
    \langle q \rangle = \left(
    \begin{array}{c}
    q_{1,2} \\
    q_{3,2}
    \end{array}
\right) = \left(
    \begin{array}{cc}
    \mu_1 \mathbf 1_{N_1}&0 \\
    0&\mu_3 \mathbf 1_{\widetilde N_2-N_1}\\
    0&0
    \end{array}
\right) \,\,\,\,
 \langle \widetilde q^T \rangle = \left(
    \begin{array}{c}
    q_{2,1} \\
    q_{2,3}
    \end{array}
    \right)= \left(
    \begin{array}{cc}
    \mu_1 \mathbf 1_{N_1}&0 \\
    0&\mu_3 \mathbf 1_{\widetilde N_2-N_1}\\
    0&0
    \end{array}
\right)
\end{equation}
we see that the global symmetry breaks as
\begin{equation}\label{glob}
    U(N_1) \times U(\widetilde N_2) \times U(N_3) \longrightarrow U(N_1)_D \times 
U(\widetilde N_2-N_1)_D \times U(N_1+N_2-\widetilde N_2)
\end{equation}
This implies that the Goldstone bosons are 
$\widetilde N_2^2 + 2(\widetilde N_2-N_1)(N_1+N_3-\widetilde N_2)$.
The first $\widetilde N_2^2$ Goldstone bosons come 
from the upper $\widetilde N_2 \times \widetilde N_2$
block matrices in the quark fields, exactly the same as in ISS. The
second part is a bit different. In fact in ISS, with equal masses,
the Goldstone bosons which come from the lower $(N_1+N_3-\widetilde N_2) 
\times \widetilde N_2$
sector in the quarks matrices, are $2 \widetilde N_2 (N_3+N_1-\widetilde N_2)$. 
In this case,
since we started with lesser flavor symmetry, there are $2 N_1 (N_3+N_1
-\widetilde N_2)$
massless Goldstone bosons fewer than in ISS. We have to control the other
directions.
From the scalar potential we have to
compute the masses that the fields acquire expanding around the
vacuum. 
The relevant expansions for
the potentially tachyonic 
directions are the ones around the vevs of the quarks
\begin{eqnarray} \label{expvev}
q_{12} &=& \left(
\begin{array}{cc}
\mu_1 +\phi_1&\phi_2 
\end{array}
\right)
\quad
q_{21}= \left(
\begin{array}{c}
\mu_1 +\widetilde \phi_1\\
\widetilde \phi_2 
\end{array}
\right)
\nonumber \\
q_{23}&=& \left(
\begin{array}{cc}
\phi_3&\mu_3 + \phi_4\\
\phi_5&\phi_6
\end{array}
\right)
\quad
q_{32}= \left(
\begin{array}{cc}
\widetilde \phi_3&\widetilde \phi_5 \\
\mu_3 + \widetilde \phi_4&\widetilde\phi_6
\end{array}
\right)
\end{eqnarray}
The relevant terms of the scalar potential
come from the $F$-terms of the mesons
\begin{equation}\label{scal}
    V =  |F_{M_{11}}|^2 + |F_{M_{13}}|^2 + |F_{M_{31}}|^2 +|F_{M_{33}}|^2 
\end{equation}
\\
If we study the mass terms of the fields $\phi_5$ and $\widetilde
\phi_5$ we note that they are not zero, since $\mu_1 \neq \mu_3$. In
fact their mass matrix is\footnote{
From now on we will consider all the mass terms as real.}
\begin{equation}\label{massphi5}
    \left(\begin{array}{cc}
      \phi_5 & \widetilde \phi_5^\dagger
    \end{array}
    \right)
    \left(
\begin{array}{cc}
  \mu_1^2 & -\mu_3^2 \\
  -\mu_3^2 & \mu_1^2
\end{array}
\right) \left(
\begin{array}{c}
  \phi_5^\dagger\\
  \widetilde \phi_5
\end{array}
\right)
\end{equation}
with eigenvalues $\mu_1^2 \pm \mu_3^2$. A
minimum of the scalar potential without tachyonic directions 
imposes a constraint on the
masses, $\mu_1>\mu_3$, 
consistent with the analysis
of ISS.

We can ask now what happens if $\mu_1<\mu_3$. The vacua we
studied before are not true vacua any longer, but they have tachyonic
directions in the quark fields. The meta-stable vacua 
are obtained
choosing 
the vevs of $q_{1,2}$ and $q_{2,1}$ to be zero, and the vevs
of the other quarks to be
\begin{equation}\label{vev222}
q_{3,2}=q_{2,3}^T
    = \left(
    \begin{array}{c}
    \mu_3 \mathbf 1_{\widetilde N_2}\\
    0
    \end{array}
    \right)
\end{equation}

The differences in the two cases
are the value of the scalar potential and the pseudo-moduli. In fact
in the first limit $V_{vac}=(N_1+N_3-\widetilde N_2)|h \mu_3^2|^2$, 
and in the second
limit the scalar potential is $V_{vac}=(N_3-\widetilde N_2)|h \mu_3^2|^2+N_1|h
\mu_1^2|^2$. Since we choose the masses to be different, but of the
same order,
both cases have long lived meta-stable vacua. As far as the pseudo-moduli are
concerned,
in the case analyzed during the paper, they come out from a block of the
$M_{3,3}$ meson, and in this case they come out from the whole $M_1$
meson and from a diagonal block 
$(N_3-\widetilde  N_2)\times(N_3-\widetilde N_2)$ of the 
$M_{3,3}$ meson.

\section{Hierarchy of scales}

One of the main approximation we used to find metastable 
vacua has been to neglect the fact that the odd nodes 
are gauge nodes. In order to treat them as flavours groups
in the region of interest, it is necessary that their gauge couplings are
lower than the couplings of the even nodes.
We can treat the odd groups as flavour groups 
only if this relation holds.  

In order to substantiate this idea we have to relate the electric scale 
of the flavour group to the other scales of the theory. 
The latter ones are the strong coupling scale of the gauge theories,
$\Lambda_{2i}$, and the supersymmetry breaking scale 
$\mu$, which is the value 
of the linear term in the dual version of the theory.

We must impose the groups related 
to the flavour/odd nodes
to be less coupled than the gauge/even groups
in the magnetic region.
A similar analysis was performed in
\cite{Forste}.

There are six possibilities, shown in Figure 1 in section \ref{BETAF}. 
We have already
discussed what happens in all these different cases. We will now 
show how to derive the formulas (\ref{condscales}) and (\ref{condscalebis}).

Let's denote by $f$ all the objects related to the flavour
group, and by $g$ all the objects related to the gauge
group.  
We have to distinguish four different cases, 
all with $\widetilde b_f > \widetilde b_g$ 
\footnote{The opposite inequality do not require this analysis, since at low
energy the flavours are always less coupled than the gauge.}. 
In fact the flavours
can be IR free or UV free in the electric description (i.e.
above the scale $\Lambda_{2i}$) and  
also UV free
or IR free in the magnetic description. 

We start studying a single case, and then we will comment about the others.
Let's study the case (2) in Figure 1, where
the flavours are 
UV free in the electric and IR free in the magnetic 
description, i.e. $b_f>0$ and $\widetilde b_f<0$.

We require that after Seiberg duality the gauge coupling $g_g$ is
larger than the flavour coupling $g_f$.
More precisely we require that this happens at the supersymmetry breaking scale
$\mu$
\be
\frac{1}{g_f^2 (\mu)} > \frac{1}{g_g^2 (\mu)}  \qquad \Rightarrow \qquad 
\widetilde b_f \log \left(\frac {\widetilde \Lambda_f}{\mu} \right)<
\widetilde b_g \log \left(\frac {\widetilde \Lambda_g}{\mu} \right)
\ee
from which follows
\be \label{disugflav}
\widetilde \Lambda_f > \left(\frac{\widetilde \Lambda_g}{\mu}\right)
^{\frac{\widetilde b_g - \widetilde b_f}{\widetilde b_f}} \widetilde \Lambda_g
>\widetilde \Lambda_g
\ee
The scale matching relation
coming from Seiberg duality 
\be
\Lambda_g^{3n_g-n_f} \widetilde \Lambda_g^{2n_f-3n_g} = \hat \Lambda_g^{n_f}
\ee
fixes $\Lambda_g = \widetilde \Lambda_g$, if we choose the intermediate 
scale to be $\hat \Lambda_g = \Lambda_g$.

For the flavour scale we observe that, at the scale $\Lambda_g$, where
we perform Seiberg duality, the coupling in the electric description
for the odd node is the same that the coupling of the magnetic
description, and this implies
\be \label{figone}
g_f = \widetilde g_f \quad \rightarrow \quad 
\left( \frac{\Lambda_f}{\Lambda_g} \right)^{b_f}=
\left( \frac{\widetilde \Lambda_f}{\Lambda_g} \right)^{\widetilde b_f}
\ee
We can now write (\ref{disugflav}) in term of the electric scales ($\Lambda_f$
and $\Lambda_g$) using (\ref{figone}), and we obtain
\be
\Lambda_f < \mu^{\frac{\widetilde b_f - \widetilde b_g}{b_f}} 
\Lambda_g^{\frac{\widetilde b_g - \widetilde b_f + b_f}{b_f}}
\ee
Since the exponent of
$\mu$ is positive we have
\be  \label{disfin}
\frac{\widetilde b_f - \widetilde b_g}{b_f}>0 \quad \rightarrow \quad 
\Lambda_f < \left(\frac{\mu}{\Lambda_g}\right)^
{\frac{\widetilde b_f - \widetilde b_g}{b_f}} \Lambda_g \ll \Lambda_g
\ee
This imposes a stronger constraint on the scale of the flavour
group $\Lambda_f$. 
In fact it is not enough to choose it lower than the gauge
strong coupling scale $\Lambda_g$. It is also constrained by (\ref{disfin}).
The next figure explains what happens
\begin{center}
\begin{tabular}{c|c}
    \includegraphics[width=7cm]{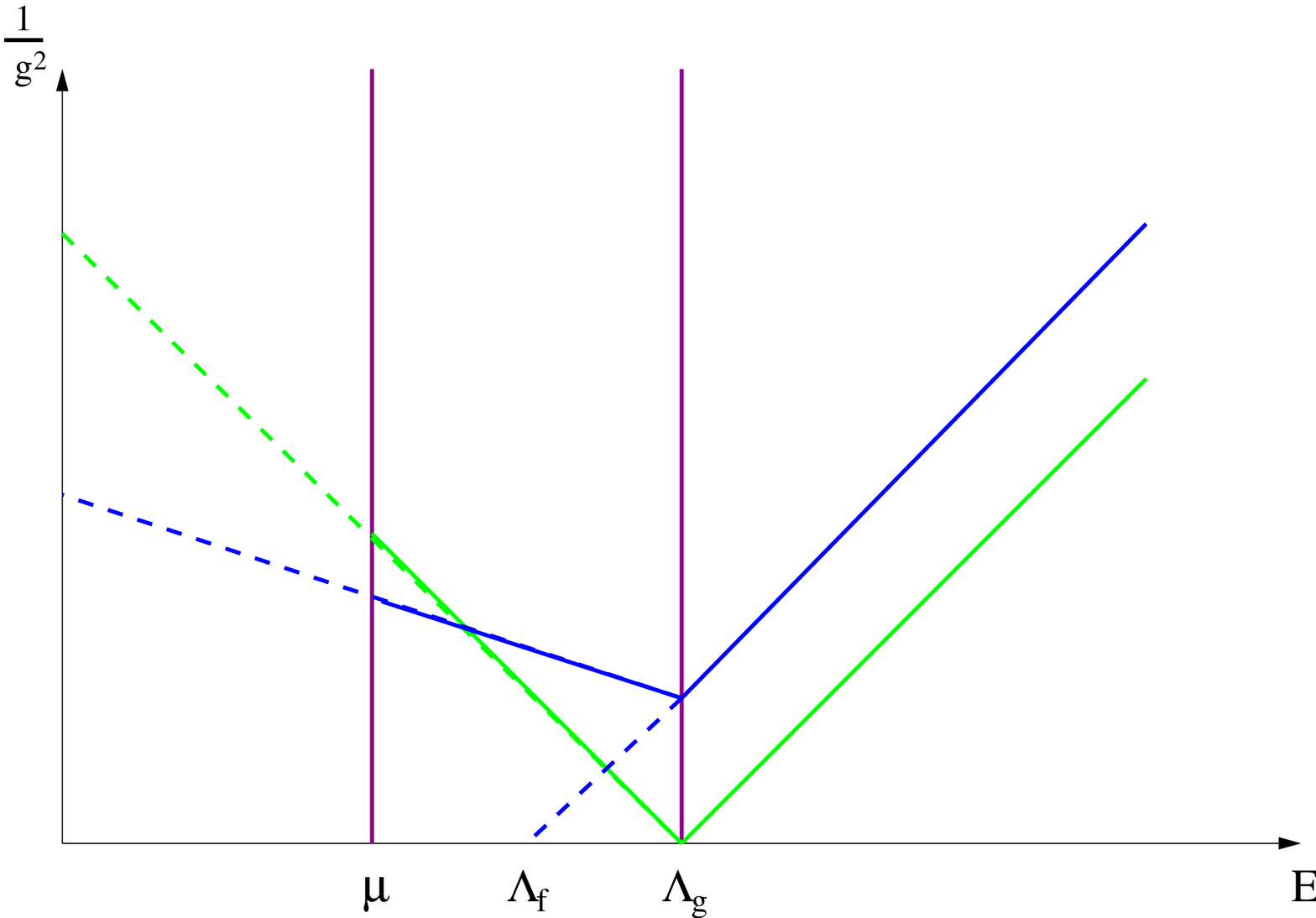}&
    \includegraphics[width=7cm]{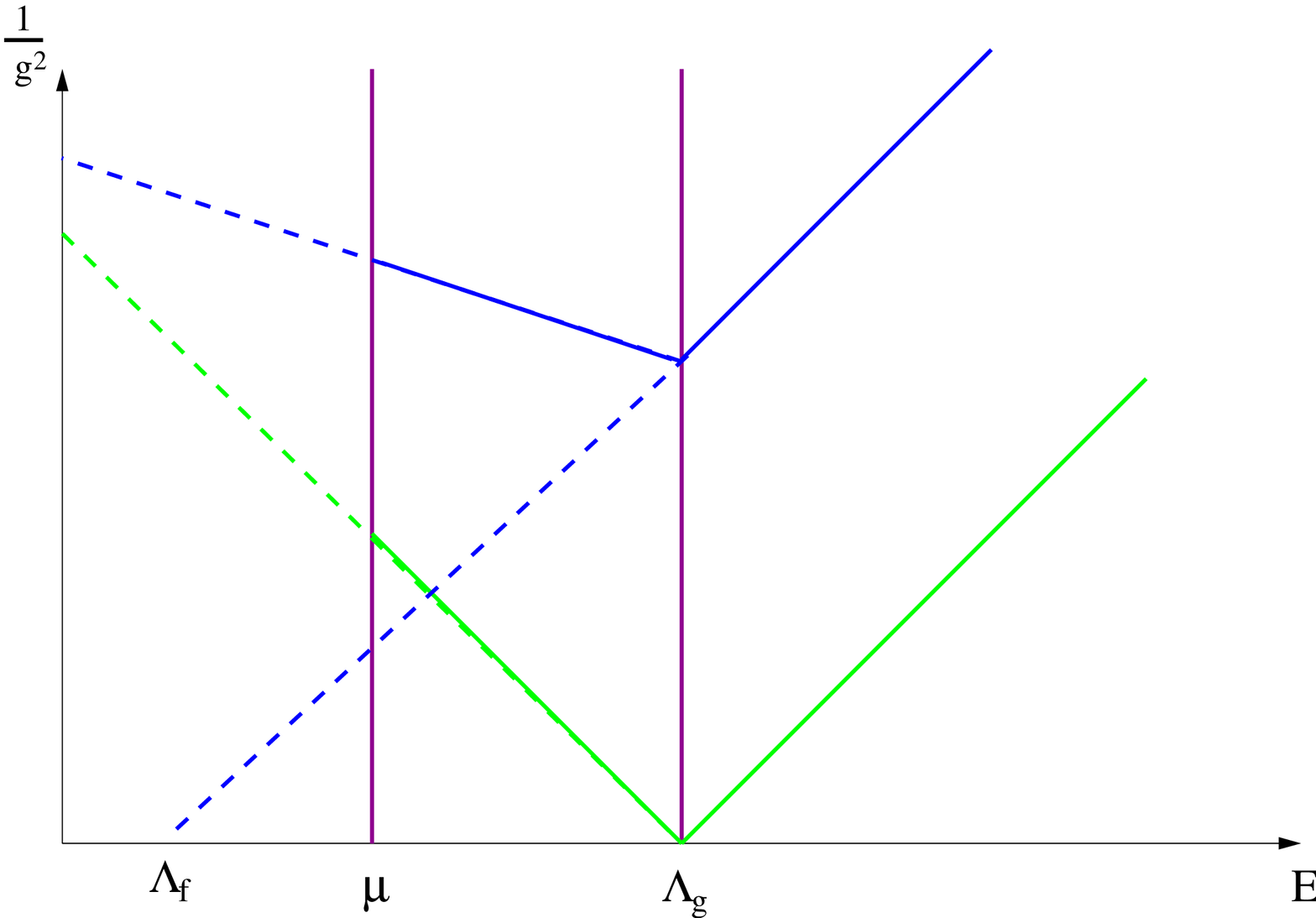}\\
\end{tabular}
\end{center}
In the first picture the scale $\Lambda_f$ is lower than 
$\Lambda_g$ but not enough: at the 
breaking scale it is not possible to neglect the contribution
coming from $\widetilde g_f$.
Instead, if
 we constrain the scale $\Lambda_f$ using (\ref{disfin}),
we obtain the runnings depicted in the second picture:
here the flavour groups are less coupled than the gauge groups at
the supersymmetry breaking scale.

As explained above there are four different possibilities.
The second possibility is that the flavours are UV free both in the
electric description and in the magnetic description,
with $\widetilde b_f > 0$.
The analysis is the same as before, and we obtain the same inequality
as (\ref{disfin}).
However this situation requires a more careful analysis, since in
the infrared the gauge coupling associated to the flavour group
develops a strong dynamics which has to be taken under control.

For the other two possibilities, where $b_f<0$, one finds 
\be \label{disfin2}
\Lambda_f > \left(\frac{\Lambda_g}{\mu} \right)^
{\frac{\widetilde b_g - \widetilde b_f}{b_f}}  
\Lambda_g \gg \Lambda_g
\ee

The general recipe we learn from this analysis can be summarized in 
three different cases

\begin{itemize} 
\item If the inequality $\widetilde b_f < \widetilde b_g$ holds one
has simply to choose $\Lambda_f \ll \Lambda_g$ or 
$\Lambda_f \gg \Lambda_g$ if $b_f>0$ or $b_f<0$ respectively
as in (\ref{SCALE1},\ref{SCALE2}).
\item If $\widetilde b_f > \widetilde b_g $ we can still distinguish two cases
\begin{itemize}
\item In the first case $b_f>0$, and we have to constraint 
$\Lambda_f$ with (\ref{disfin}).
\item In the second case $b_f<0$, and we have to constraint 
$\Lambda_f$ with (\ref{disfin2}).
\end{itemize}
\end{itemize}

\section{$A_5$ classification}\label{acinque}
We study 
$A_5$ quiver gauge theories obtained gluing all the possible
combinations of $A_3$ which present metastable vacua,
i.e. the one of section (\ref{metA3})

We analyze the beta function coefficients for 
these $A_5$ quiver gauge theories, 
with gauge group $U(N_1)\times U(N_2)\times U(N_3)\times
U(N_4)\times U(N_5)$.
The even
nodes are in the IR free window 
\be
\label{seib10}
N_2 < N_1+N_3< \frac{3}{2} N_2 \qquad N_4 < N_3+N_5< \frac{3}{2} N_4
\ee 
We write in the table the beta coefficients of the third node of the
$A_5$, specifying the range, compatible with (\ref{seib10}), when this
node is UV free or IR free in the electric and in the magnetic
descriptions, respectively.
The  
table classifies the possible
$A_5$ quiver gauge theories 
which present alternate Seiberg 
dualities and which have 
metastable vacua.


As explained in section \ref{extension}
we can obtain an $A_n$ quiver gauge theory by
gluing the $A_3$ patches.
For the renormalization group,
the internal flavour nodes 
of the $A_n$ chain behave as
the third node of the $A_5$ patches.

The table does not say anything about the external nodes of the $A_n$.
In the electric theory one has $b_1 = 3 N_1 -N_2$
and $b_n = 3N_n-N_{n-1}$; after duality, in the low
energy description we have $\widetilde b_1 = N_1 + N_2 - N_3$,
and $\widetilde b_n = N_n + N_{n-1} - 2N_{n-2}$.
The possible values for $\widetilde b_1$
and $\widetilde b_n$ have to be studied separately.

%
\begin{landscape}
\begin{center}
\begin{tabular}{c|c|c|c|c}
Ranks of $A_5$
&
$\begin{array}{c}
\text{Further}\\ 
\text{condition} (I)
\end{array}$
&
$\begin{array}{c}
\text{Further}\\ 
\text{condition} (II)
\end{array}$
&
$\begin{array}{c}
\text{electric}\\
b-\text{factor}
\end{array}$
&
$\begin{array}{c}
\text{magnetic}\\
b-\text{factor}
\end{array}$
\\
\hline
$\begin{array}{c}
N_1 < N_2 \leq  N_3 < N_4 \leq N_5 \\ 
\end{array}$
&
&
$\begin{array}{c} 
 N_2+N_4<3 N_3  \\
\\
3 N_3< N_2+N_4                
\end{array}$
&
$\begin{array}{c}
b_3>0\\
\\
b_3<0
\end{array}$
&
$\begin{array}{c}
\widetilde b_3<0 \\
\\
\widetilde b_3<0 
\end{array}$
\\
\hline
$N_1 < N_2 > N_3 < N_4 \leq N_5$ 
&
&
& $b_3<0$ & $\widetilde b_3<0$ \\
\hline
$N_1 \geq N_2 > N_3 < N_4 \leq N_5$  
&&& $b_3<0$ & $\widetilde b_3<0$  \\
\hline
$\begin{array}{c}
N_1<N_2>N_3<N_4>N_5 \\
\end{array}$
&
$\begin{array}{c}
N_3<N_1+N_5\\
\\
\\
N_3>N_1+N_5\\
\end{array}$
&
$\begin{array}{c}
N_2+N_4<3 N_3\\  
\\
3 N_3< N_2+N_4\\  
\hline
N_2+N_4< N_3 + 2 N_1 + 2 N_5\\
\\
N_3 + 2N_1 + 2N_5 < N_2+N_4
\end{array}
$
&
$\begin{array}{c}
b_3>0\\
\\
b_3<0\\
\hline
b_3>0\\
\\
b_3>0
\end{array}$
&
$\begin{array}{c}
\widetilde b_3<0\\
\\
\widetilde b_3<0\\
\hline
\widetilde b_3<0\\
\\
\widetilde b_3>0
\end{array}$
\\
\hline
$\begin{array}{c}
N_1<N_2\leq N_3\geq N_4>N_5 \\
\end{array}$ 
&&
$\begin{array}{c}
N_2+N_4< N_3+2 N_1+2 N_5\\
\\
 N_3+2 N_1+2 N_5<  N_2+N_4
\end{array}$
&
$\begin{array}{c}
b_3>0\\
\\
b_3>0
\end{array}$
&
$\begin{array}{c}
\widetilde b_3<0\\
\\
\widetilde b_3>0
\end{array}$
\\
\hline
$\begin{array}{c}
N_1<N_2\leq N_3<N_4>N_5 \\
\end{array}$
&
$\begin{array}{c}
N_3<N_1+N_5\\
\\
\\
N_3>N_1+N_5\\
\end{array}$
&
$\begin{array}{c}
 N_2+N_4<3 N_3\\
\\
3 N_3< N_2+N_4\\
\hline
N_2+N_4< N_3+2 N_1+2 N_5\\
\\
N_3+2 N_1+2 N_5<  N_2+N_4
\end{array}$
&
$\begin{array}{c}
b_3>0\\
\\
b_3<0\\
\hline
b_3>0\\
\\
b_3>0
\end{array}$
&
$\begin{array}{c}
\widetilde b_3<0\\
\\
\widetilde b_3<0\\
\hline
\widetilde b_3<0\\
\\
\widetilde b_3>0
\end{array}$
\end{tabular}
\end{center}
\small{In the first column we report all the possible inequalities among the 
$A_5$ rank numbers
consistent with (\ref{seib10}).
Moving from left to right the further condition fix the signs of $b_3$,
$\widetilde b_3$}.
\end{landscape}

\newpage

\end{document}